\documentclass[twoside,twocolumn,9pt]{article}
\usepackage{extsizes}
\usepackage[super,sort&compress,comma]{natbib} 
\usepackage[version=3]{mhchem}
\usepackage[left=1.5cm, right=1.5cm, top=1.785cm, bottom=2.0cm]{geometry}
\usepackage{balance}
\usepackage{mathptmx}
\usepackage{sectsty}
\usepackage{graphicx} 
\usepackage{lastpage}
\usepackage[format=plain,justification=justified,singlelinecheck=false,font={stretch=1.125,small,sf},labelfont=bf,labelsep=space]{caption}
\usepackage{float}
\usepackage{fancyhdr}
\usepackage{fnpos}
\usepackage{amssymb}
\usepackage[english]{babel}
\addto{\captionsenglish}{%
  
}
\usepackage{array}
\usepackage{droidsans}
\usepackage{charter}
\usepackage[T1]{fontenc}
\usepackage[usenames,dvipsnames]{xcolor}
\usepackage{setspace}
\usepackage[compact]{titlesec}
\usepackage{hyperref}
\usepackage{dblfloatfix}
\usepackage{epstopdf}
\usepackage{float}
\definecolor{cream}{RGB}{222,217,201}

\begin{document}

\pagestyle{fancy}
\thispagestyle{plain}
\fancypagestyle{plain}{
\renewcommand{\headrulewidth}{0pt}
}

\makeFNbottom
\makeatletter
\renewcommand\LARGE{\@setfontsize\LARGE{15pt}{17}}
\renewcommand\Large{\@setfontsize\Large{12pt}{14}}
\renewcommand\large{\@setfontsize\large{10pt}{12}}
\renewcommand\footnotesize{\@setfontsize\footnotesize{7pt}{10}}
\makeatother

\renewcommand{\thefootnote}{\fnsymbol{footnote}}
\renewcommand\footnoterule{\vspace*{1pt}%
\color{cream}\hrule width 3.5in height 0.4pt \color{black}\vspace*{5pt}} 
\setcounter{secnumdepth}{5}

\makeatletter 
\renewcommand\@biblabel[1]{#1}            
\renewcommand\@makefntext[1]%
{\noindent\makebox[0pt][r]{\@thefnmark\,}#1}
\makeatother 
\renewcommand{\figurename}{\small{Fig.}~}
\sectionfont{\sffamily\Large}
\subsectionfont{\normalsize}
\subsubsectionfont{\bf}
\setstretch{1.125} 
\setlength{\skip\footins}{0.8cm}
\setlength{\footnotesep}{0.25cm}
\setlength{\jot}{10pt}
\titlespacing*{\section}{0pt}{4pt}{4pt}
\titlespacing*{\subsection}{0pt}{15pt}{1pt}

\fancyfoot{}
\fancyfoot[LO,RE]{\vspace{-7.1pt}\includegraphics[height=9pt]{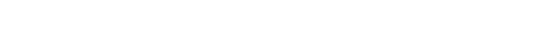}}
\fancyfoot[CO]{\vspace{-7.1pt}\hspace{13.2cm}\includegraphics{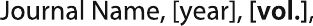}}
\fancyfoot[CE]{\vspace{-7.2pt}\hspace{-14.2cm}\includegraphics{RF}}
\fancyfoot[RO]{\footnotesize{\sffamily{1--\pageref{LastPage} ~\textbar  \hspace{2pt}\thepage}}}
\fancyfoot[LE]{\footnotesize{\sffamily{\thepage~\textbar\hspace{3.45cm} 1--\pageref{LastPage}}}}
\fancyhead{}
\renewcommand{\headrulewidth}{0pt} 
\renewcommand{\footrulewidth}{0pt}
\setlength{\arrayrulewidth}{1pt}
\setlength{\columnsep}{6.5mm}
\setlength\bibsep{1pt}

\makeatletter 
\newlength{\figrulesep} 
\setlength{\figrulesep}{0.5\textfloatsep} 

\newcommand{\topfigrule}{\vspace*{-1pt}%
\noindent{\color{cream}\rule[-\figrulesep]{\columnwidth}{1.5pt}} }

\newcommand{\botfigrule}{\vspace*{-2pt}%
\noindent{\color{cream}\rule[\figrulesep]{\columnwidth}{1.5pt}} }

\newcommand{\dblfigrule}{\vspace*{-1pt}%
\noindent{\color{cream}\rule[-\figrulesep]{\textwidth}{1.5pt}} }

\makeatother

\twocolumn[
  \begin{@twocolumnfalse}
{\includegraphics[height=30pt]{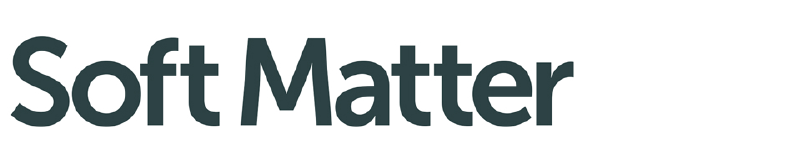}\hfill\raisebox{0pt}[0pt][0pt]{\includegraphics[height=55pt]{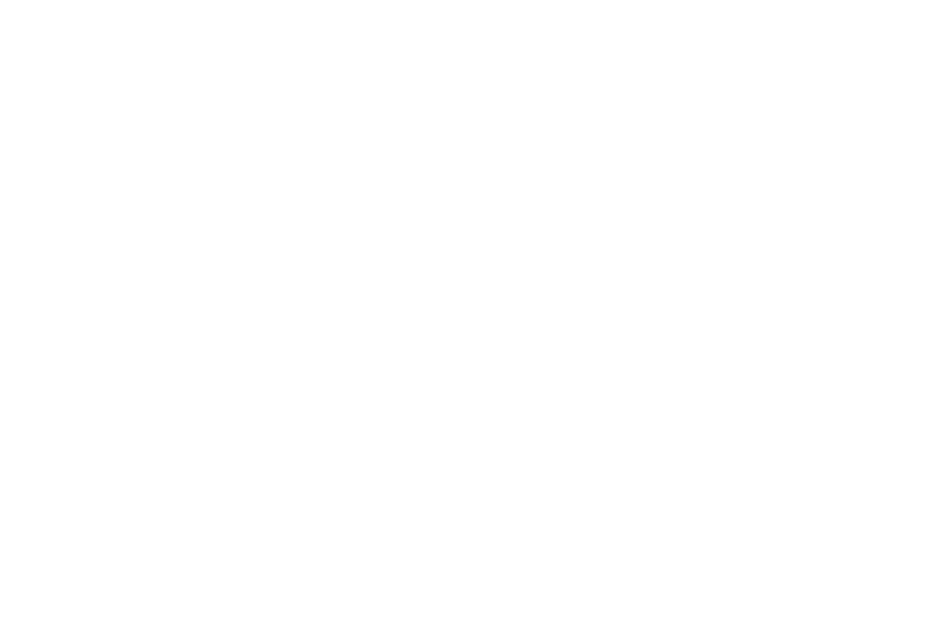}}\\[1ex]
\includegraphics[width=18.5cm]{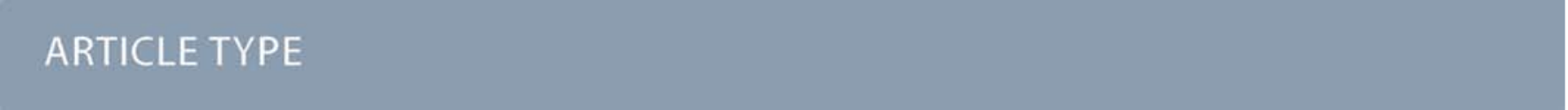}}\par
\vspace{1em}
\sffamily
\begin{tabular}{m{4.5cm} p{13.5cm} }

\includegraphics{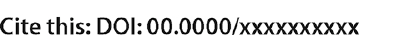} & \noindent\LARGE{\textbf{Effect of ring stiffness and ambient pressure on the dynamical slowdown in ring polymers $^\dag$}} \\
\vspace{0.3cm} & \vspace{0.3cm} \\

 & \noindent\large{Projesh Kumar Roy,\textit{$^{a,b}$} Pinaki Chaudhuri,\textit{$^{a,b, \ast}$} and Satyavani Vemparala\textit{$^{a,b, \ast}$}} \\

\includegraphics{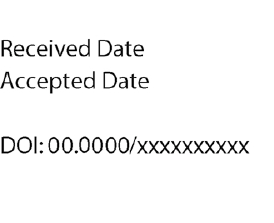} & \noindent\normalsize{Using extensive molecular dynamics simulations, we investigate the slowing down of dynamics in a 3D system of ring polymers  by varying the ambient pressure and the stiffness of the rings. Our study demonstrates that the stiffness of the rings determines the dynamics of the ring polymers, leading to glassiness at lower pressures for stiffer rings. The threading of the ring polymers, a unique feature that emerges only due to the topological nature of such polymers in three dimensions, is shown to be the determinant feature of dynamical slowing down, albeit only in a certain stiffness range. Our results suggest a possible framework of exploring the  phase space spanned by ring stiffness and pressure to obtain spontaneously emerging topologically constrained polymer glasses.} \\

\end{tabular}

 \end{@twocolumnfalse} \vspace{0.6cm}

]

\renewcommand*\rmdefault{bch}\normalfont\upshape
\rmfamily
\section*{}
\vspace{-1cm}

\footnotetext{\textit{$^{a}$~The Institute of Mathematical Sciences, C.I.T. Campus, Taramani, Chennai 600113, India; Tel: +914422543293/+914422543257; E-mail: pinakic@imsc.res.in/vani@imsc.res.in}}

\footnotetext{\textit{$^{b}$~Homi Bhabha National Institute, Training School Complex, Anushakti Nagar, Mumbai 400094, India}}


\footnotetext{\dag~Electronic Supplementary Information (ESI) available:  See DOI: 10.1039/cXsm00000x/}




\section{Introduction} 

Entanglement in dense polymeric systems can arise due to many reasons and can lead to interesting and complex rheological phenomena. For linear polymer systems, entanglement can arise from spatial confinement via the presence of other linear polymers in close contact. It is intuitive that behavior of a single polymer in dilute vs dense conditions is very different.  In the former case, the dynamics that govern the polymer behavior are largely determined by the interconnectivity of the polymer and the nature of the solvent surrounding it, while in the latter, the presence of other polymer molecules within close vicinity can significantly influence the dynamics of the polymers. Theoretical models exist to describe these two different scenarios: Rouse-Zimm model~\cite{rouse1953theory, zimm1956dynamics, de1976dynamics, zwanzig1974theoretical} for extreme dilute systems, reptation or tube models~\cite{doi1988theory,de1971reptation,doi1978dynamics,granek1997semi} for the dense linear polymer systems. 

Entanglement can also arise from topological constraints via connecting the ends of the linear polymers resulting in ring polymers and consequently exhibiting different rheological properties~\cite{Rosa_Everaers_PhysRevLett_2014, Halverson_Kremer_PhysRevLett_2012, Halverson_Kremer_JChemPhys_2011, Halverson_Kremer_JChemPhys_2011_2, rubinstein1986dynamics, cates1986conjectures, lee2015segregated, brown2001influence, suzuki2009dimension, Lee_Jung_Polymers_2019, deguchi2017statistical}.The ring polymers have been considered as model systems for understanding the compaction and organization of biologically relevant polymers like chromatin~\cite{tark2011chromatin, zhou2019effect, florescu2016large, halverson2014melt}. In addition, recent studies~\cite{mukhopadhyay2011packings, makse2000packing, batista2010crystallization, boromand2018jamming} have shown that systems composed of deformable particles such as colloids exhibit glassy behavior as a function of packing fraction. {The introduction of deformability and softness in the hard-sphere particles can induce novel packing behavior at high densities.} The ring polymer systems thus can be an excellent model platform to mimic such colloids, with the possibility of soft and tuneable interactions, to understand emergence of glassy behavior in such colloidal systems. The complex behavior of the dynamics of the ring polymer                          systems poses an interesting challenge to the theoretical understanding of polymer entanglement. In the last decade, several attempts were made to understand their dynamical as well as statistical properties using computer simulations~\cite{Halverson_Kremer_JChemPhys_2011,Halverson_Kremer_JChemPhys_2011_2, Halverson_Kremer_PhysRevLett_2012}. {Simulations of ring polymers show a sluggish decay in the stress relaxation modulus (G) at longer time-scales when compared with theoretical models for such polymers~\cite{Kapnistos_Rubinstein_Nature_2008, Pasquino_Vlassopoulos_Macrolett_2013,Tsalikis_Vlassopoulos_MacroLett_2016}.} It was postulated that such deviation is caused by ring-ring penetration, or \textit{threading} effect, which was not incorporated in the theoretical models like self-similar lattice-animal model~\cite{Obukhov_Duke_PhysRevLett_1994, Obukov_Colby_Macromol_1994} annealed-tree model~\cite{Grosberg_SoftMatter_2014, Smrek_Grosberg_JPhysCondMatt_2015}, or fractal loopy globule model~\cite{Ge_Rubinstein_Macromol_2016}. Despite pure linear polymer dense systems not exhibiting similar dynamical behavior, it was shown that the presence of even a small amount of linear polymers in the ring polymer systems can dramatically increase the relaxation times and make the ring polymer systems more sluggish~\cite{Kapnistos_Rubinstein_Nature_2008, Tsalikis_Marvrantzas_ACSMacroLett_2014, Tsalikis_Richter_Macromol_2017, zhou2019effect, Tsalikis_Marvrantzas_Macromol_2020, Tsalikis_Marvrantzas_MacromolTheoSim_2020}. Assuming a similarity of the conformations between the ring and linear polymers at a length scale smaller than the radius of gyration (R$_{g}$), some modified theories have been proposed to incorporate the threading effect in pure ring polymer systems, e.g. Tsalikis \textit{et al}~\cite{Tsalikis_Vlassopoulos_MacroLett_2016},  Dell \textit{et al}~\cite{Dell_Schweizer_SoftMatter_2018}. 

Recently, it was shown that the dynamics of the ring polymer systems mimic a glassy behavior when a randomly chosen minute fraction of the rings are pinned in the simulation box~\cite{Michieletto_Turner_PNAS_2016, Michieletto_Rosa_PhysRevLett_2017, michieletto2017ring}.  Further, it was estimated that at a large chain-length of $\sim 3500$, glass formation of the ring polymer system becomes spontaneous and independent of any external perturbations like pinning~\cite{Michieletto_Turner_PNAS_2016}. This type of glass formation is unique as it does not require any cooling, jamming, or modification of  the interaction energies~\cite{Michieletto_Turner_PhysWorld_2014}. Such glass-forming systems are termed \textit{topological-glass}~\cite{Lo_Turner_EPL_2013} which linear polymers fail to exhibit~\cite{Lee_Jung_MacromolRComm_2015}. The phenomenon of topological glass formation as well as the slow relaxation of G, was explained using the threading effect. This theory suggests that the glassy phase originates as a result of the long-range entanglements of the ring polymers in the form of threading, which creates insurmountable energetic barriers for the reptation movement of the ring polymers at larger timescale~\cite{Obukhov_Duke_PhysRevLett_1994, Obukov_Colby_Macromol_1994}. Formation of such barriers is controlled by the topology (or conformations) and entropically driven mixing of the rings. {Several studies have focused on understanding the effects of threading on the dynamics of the pure ring polymer systems  using Monte-Carlo (MC)~\cite{Lo_Turner_EPL_2013, Lee_Jung_Polymers_2019} and Molecular-Dynamics (MD)~\cite{Lee_Jung_MacromolRComm_2015, Tsalikis_Vlassopoulos_MacroLett_2016, Smrek_Grosberg_MacroLett_2016, Smrek_Rosa_MacroLett_2019} methods and experiments~\cite{OConnor_Grest_PhysRevLett_2020, Gomez_Poschel_PNAS_2020, Soh_Doyle_PhysRevLett_2019}. The theoretical studies were performed on either flexible ring polymer systems, or lattice models. Threading is a topological phenomenon, which suggests that it should be correlated with the shape of the ring polymers. However, a proper understanding of the correlation between shape and threading in ring polymers is still missing in the literature. Recent studies have started focusing on the effect of stiffness on threading parameter for systems of knotted and unknotted polymers~\cite{Guo_Zhang_Polymers_2020}.} { However, there has not been much attention on dense assemblies of ring polymers where there can be variations in the shape of the  polymers, and thereby affect threading and consequently the dynamics under equilibrium conditions.}

In this paper, we focus on emergence of dynamical slowing down in systems of equilibrated ring polymer systems and identify the domain in phase space spanned by pressure and ring stiffness, i.e. (P-$K_{\theta}$), in which the glassy regime would emerge. Importantly, we observe that stiffer ring polymers exhibit dynamical slowdown at smaller ambient pressures. We explicitly calculate the total effect of threading in such systems, and show that the threading effect is directly contributing to the observed slowdown at certain P-$K_{\theta}$ state points. The pressure induced conformational changes of the ring polymers and their effects on the threading properties in equilibrium are also discussed.

The paper is organized as follows. After this introductory section, we provide a section on methods where the model system that we study and the details of the simulations that we are perform are outlined. Subsequently, there is an extensive section discussing our results, where we detail the dynamical and structural properties of the system, and the correlations between the two. Finally, we have a concluding section where we summarize our findings and provide some future perspectives.

\section{Methods}

\subsection{Simulation details}

\begin{figure}[h]
\centering
\includegraphics[scale=0.5]{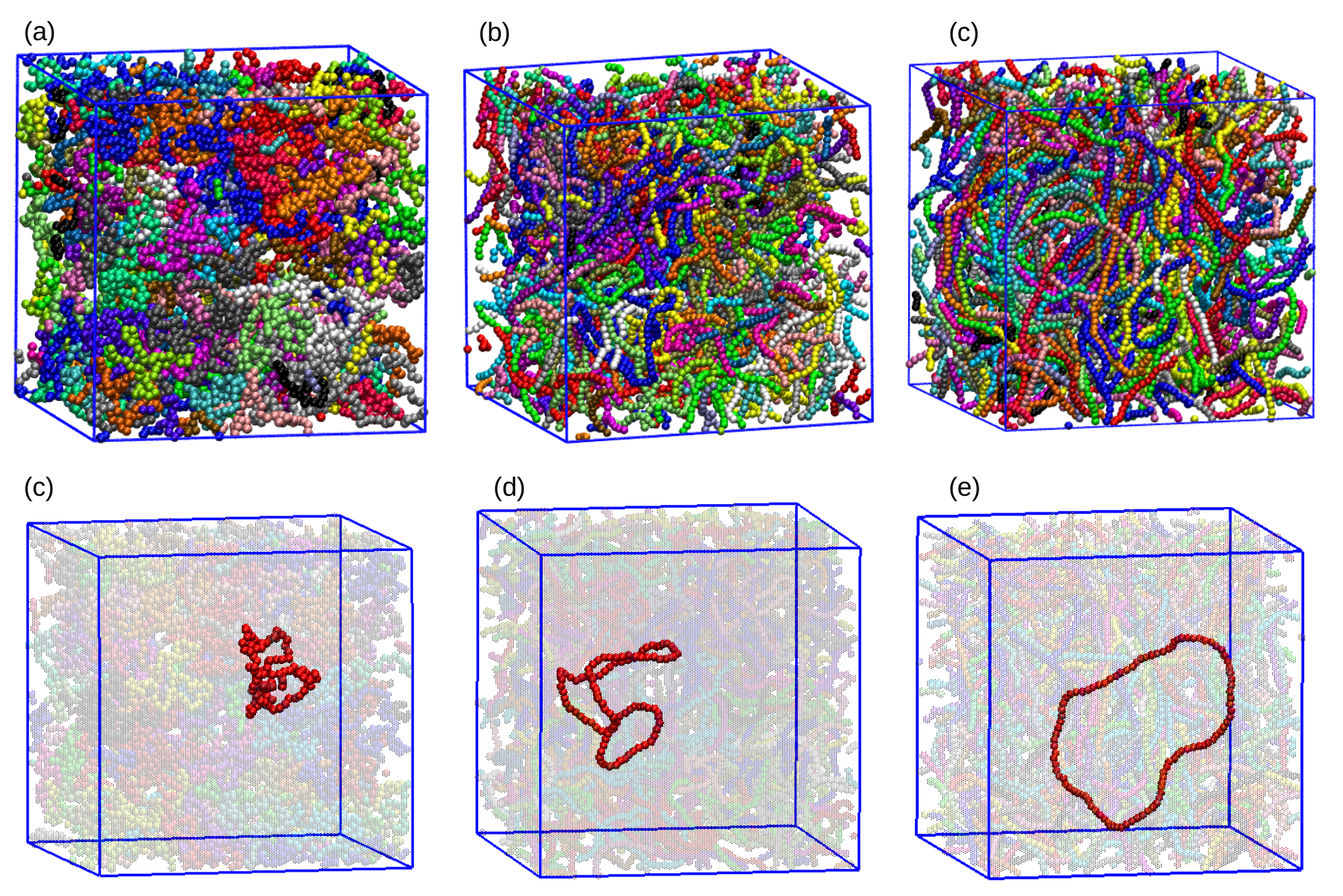}
\caption{(Top) Snapshots of equilibrium configurations sampled at a pressure value of $P=0.010$ and different ring stiffness, viz. $K_{\theta}$=1 (a),10 (b), and 20 (c). (Bottom) Conformation of a single polymer inside the bulk at these P and $K_{\theta}$ values in (c),(d), and (e).}
\label{fig:snapshot}
\end{figure}

\begin{figure*}
\includegraphics[scale=1]{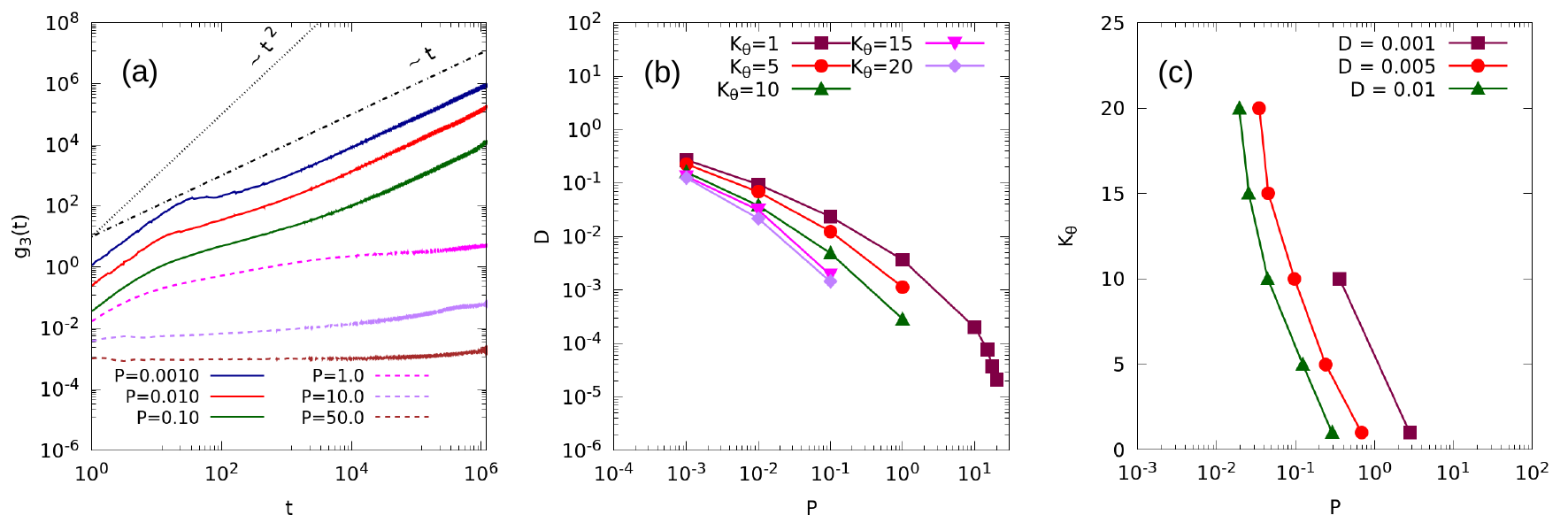}
\caption{{(a) Temporal variation of the mean square displacement of the center-of-mass of ring polymers ($g_3(t)$) for $K_{\theta}$=20 and different pressures ($P$) as marked. Data are shown for regimes where equilibrium has been attained (in solid lines) and where the dynamics is out-of-equilibrium within time-window of observation (in dashed lines). Data is shown for $t \geq t_d$. (b) Variation of center-of-mass diffusion constants (D) with pressure, calculated from the equilibrium $g_3(t)$ data  for different values of  $K_{\theta}$ as indicated. (c) Estimated iso-diffusivity contours in $P-K_{\theta}$ plane, using data shown in (b), for different diffusion constant values, as marked.}} 
\label{fig:diffusion}
\end{figure*}

A series of molecular dynamics (MD) simulations of ring polymer systems are performed to probe the slowing down in dynamical behavior of such systems with different values of ring stiffness and applied pressure. The pairwise potential of the ring polymers is modeled using the Kremer-Grest (KG) {bead-spring} model~\cite{Kurt_Grest_JChemPhys_1990, Halverson_Kremer_JChemPhys_2011, Halverson_Kremer_JChemPhys_2011_2}. In KG model, the repulsive part of a truncated and shifted Lennard-Jones potential,  aka Weeks-Chandler-Anderson (WCA) potential~\cite{Weeks_Andersen_JChemPhys_1971}, is used to model the monomer-monomer non-bonded interactions:

\begin{equation}
U_{\text{pair}}(r_{ij}) = \begin{cases}
4\epsilon \left [ \left (\frac{\sigma}{r_{ij}} \right )^{12} - \left (\frac{\sigma}{r_{ij}} \right )^{6} + \frac{1}{4} \right ] & r_{ij} \leq 2^{1/6}\sigma \\
0 & r_{ij} > 2^{1/6}\sigma 
\end{cases}
\label{eqn:WCA}
\end{equation}

where $\sigma$ and $\epsilon$ are length and energy scales, respectively. $\epsilon$ is measured in $k_BT$ units, where $k_B$ is the Boltzmann constant, and $T$ is the temperature. {$r_{ij}$ defines the distance between any two monomers in the system.} A finitely extensible non-linear elastic (FENE) potential~\cite{Kurt_Grest_JChemPhys_1990} is used to define the covalent bonds between the monomers:

\begin{equation}
U_{\text{FENE}}(d_{ij}) = \begin{cases}
-0.5 \kappa R_0^2ln \left [1 - \left ( \frac{d_{ij}}{R_0} \right )^2 \right ] & d_{ij} < R_0 \\
\infty & d_{ij} \geq R_0
\end{cases}
\label{eqn:FENE}
\end{equation}

where $\kappa = 30$, $R_0 = 1.5$. {Here, $d_{ij}$ denotes the distance between two bonded monomers in a ring polymer.} In our paper, we have used the Kratky-Porod model~\cite{Doi_Edwards}, recently used in reference~\citenum{Guo_Zhang_Polymers_2020}, for controlling the ring stiffness,

\begin{equation}
U_{\text{angle}}(\theta_i) = K_{\theta}[1 - cos(\theta_i - \pi)]
\label{eqn:angle}
\end{equation}

where $K_{\theta}$ is the stiffness parameter, which is varied in our simulations. {$\theta_i$ defines the angle between three successive monomers centered at $i^{th}$ monomer in a ring polymer.} The temperature of the system is kept fixed at 1.0 for all simulations. The mass of each monomer was taken to be 1.0. 

In this paper, we have simulated a ring polymer system of 100 rings, each with 100 monomers (total 10000 particles). {The ring-size was kept smaller than used in the previous simulations~\cite{Michieletto_Rosa_PhysRevLett_2017, Michieletto_Turner_PNAS_2016}, to understand if threading phenomenon can be envisaged for small ring sizes as well.} We randomly packed the ring polymers in a cubic box of length 10$^3$ (in units of $\sigma$) with periodic boundary conditions to create the initial low-density systems. Further, we made sure that the rings remain non-concatenated during packing. The Nos{\'e} - Hoover chains thermostat~\cite{Nose_JChemPhys_1984, Martyna_Tuckerman_JChemPhys_1992} and barostat~\cite{Nose_Klein_MolPhys_1983} were used to control the temperature and the pressure of the system, respectively. We use an overdamped barostat, which couples to the system over a {timescale of $t_d=1.0$}.  The molecular dynamics simulations were done using LAMMPS~\cite{LAMMPS}.

The pressure on the initial low density systems was iteratively increased for 100 steps with 1000 step intermediate relaxation periods using a time-step of 0.003 until the target pressure is reached. After the compression cycle is completed, the systems were allowed to relax for $\sim 4 \times 10^7$ steps. A total of 10$^8$ production steps were subsequently ran with a larger time-step of 0.012. This procedure was continued for different state-points from $K_{\theta} = 1$ to $20$, and $P = 0.001$ to $20.0$. A few snapshots of the simulation box and an arbitrarily chosen ring polymer at the end of the production runs are included in Figure~\ref{fig:snapshot}.

\section{Results}

\subsection{Equilibrium dynamics}

To monitor the dynamical behavior of the ring polymer systems, we study the time evolution of the mean square deviation of the center-of-mass displacement of the ring polymers, $g_3(t)$, which is defined as,
\begin{equation}
	g_3(t) = \langle[r_{\text{CM}}(t_0) - r_{\text{CM}}(t_0+t)]^2\rangle
	\label{eqn:msd} 
\end{equation}
where $t_0$ is the reference time origin and  $\langle\rangle$ denotes the average over all such time origins and all rings. The corresponding analysis is discussed via Figure~\ref{fig:diffusion}.

{In Figure~\ref{fig:diffusion}(a), we show the $g_3(t)$ data for a fixed value of ring stiffness, viz. $K_\theta=20$, i.e. for the case where the rings are relatively stiffer. We observe that for a range of pressure values, the long-time behavior is linear, i.e.  $g_3(t) \sim t$ (marked in solid lines). This implies that the dynamics of the rings is diffusive  and therefore indicates that the equilibrium regime is being sampled for these ambient pressures. At larger pressure values ($P\ge1.0$), this diffusive regime is not attained within the time window of our measurement. Rather, with increasing pressure, a prolonged plateau-like regime is observed (see data shown in dashed lines in Figure~\ref{fig:diffusion}(a)), which would correspond to the caging of a ring by neighboring rings. The observed caging behavior is reminiscent of the dynamics typical to glass-forming liquids on the approach to dynamical arrest \cite{binder2011glassy}. With increasing pressure and thus increased density, the localization of the rings significantly increases, as is evident from the decreasing height of the plateau-like regime in $g_3(t)$. Such dynamical behavior is also observed in the rings with other $K_\theta$ values. In Supplementary Information (Figure SI.3(a)), we show data for $K_\theta=1$ as an additional example.} 

{We note here that variation in the short-time dynamics is observed with changing pressure at fixed $K_\theta$, even at constant temperature. This occurs due to varying box fluctuations resulting from the coupling to a barostat. However, and most importantly, we have verified that this has no impact on the diffusive dynamics observed at long times (see Supplementary Information Figure SI.2(a)), which is occurring much later vis-a-vis the barostat's coupling timescale $t_d$, in the range of pressures where dynamical slowing down is happening. Relatedly, in all cases, we now display dynamical data for timescales larger than the barostat coupling timescale $t_d$.} 

It is evident in Figure~\ref{fig:diffusion}(a) that the long-time diffusive dynamics shows a slowing down with increasing pressure. This is illustrated via the plot showing the dependence of the extracted center-of-mass diffusion coefficient $D=\lim\limits_{t\to\infty}g_3(t)/6t$ versus pressure ($P$) as shown in Figure~\ref{fig:diffusion}(b), for a range of $K_{\theta}$ values. The data clearly suggests that the onset of slow dynamics depends on both ring polymer stiffness, $K_{\theta}$, and the pressure: rings with larger $K_{\theta}$ values clearly display {that the eventual onset of glassiness (i.e. $D=0$ or dynamical arrest) would occur at lower pressures if one were to extrapolate the $D$ vs $P$ curves}. This effect can be seen more clearly  in Figure~\ref{fig:diffusion}(c), where we plot the isodiffusivity lines corresponding to $D=0.01, 0.005, 0.001$ in the $K_\theta, P$ plane. {These iso-diffusivity lines  act as guides to the shape of the eventual line of glass transition in this phase-plane as we vary the stiffness of the ring polymers and the ambient pressure.}

\subsection{Structural Characterization}

\begin{figure}[]
\centering
\includegraphics[scale=0.55]{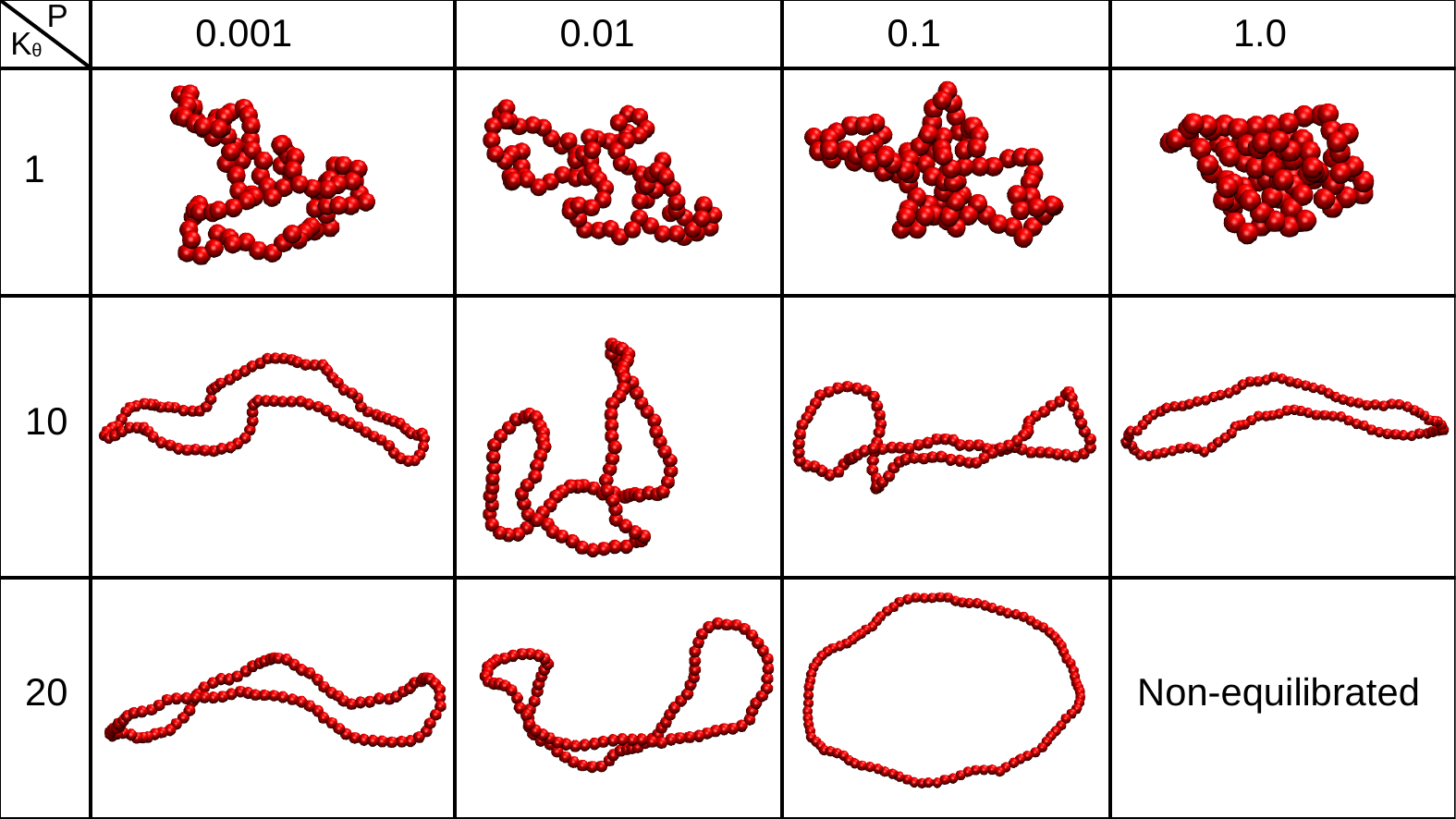}
\caption{Structures of an arbitrarily chosen ring polymer at various $K_{\theta}$ and pressures are shown in tabular form. All structures shown here belong to the diffusive regime except for $K_{\theta}$=20 and P=1.0, where we could not find a sample of an equilibrated trajectory.}
\label{fig:single_molecule}
\end{figure}

\begin{figure}
\centering
\includegraphics[scale=0.75]{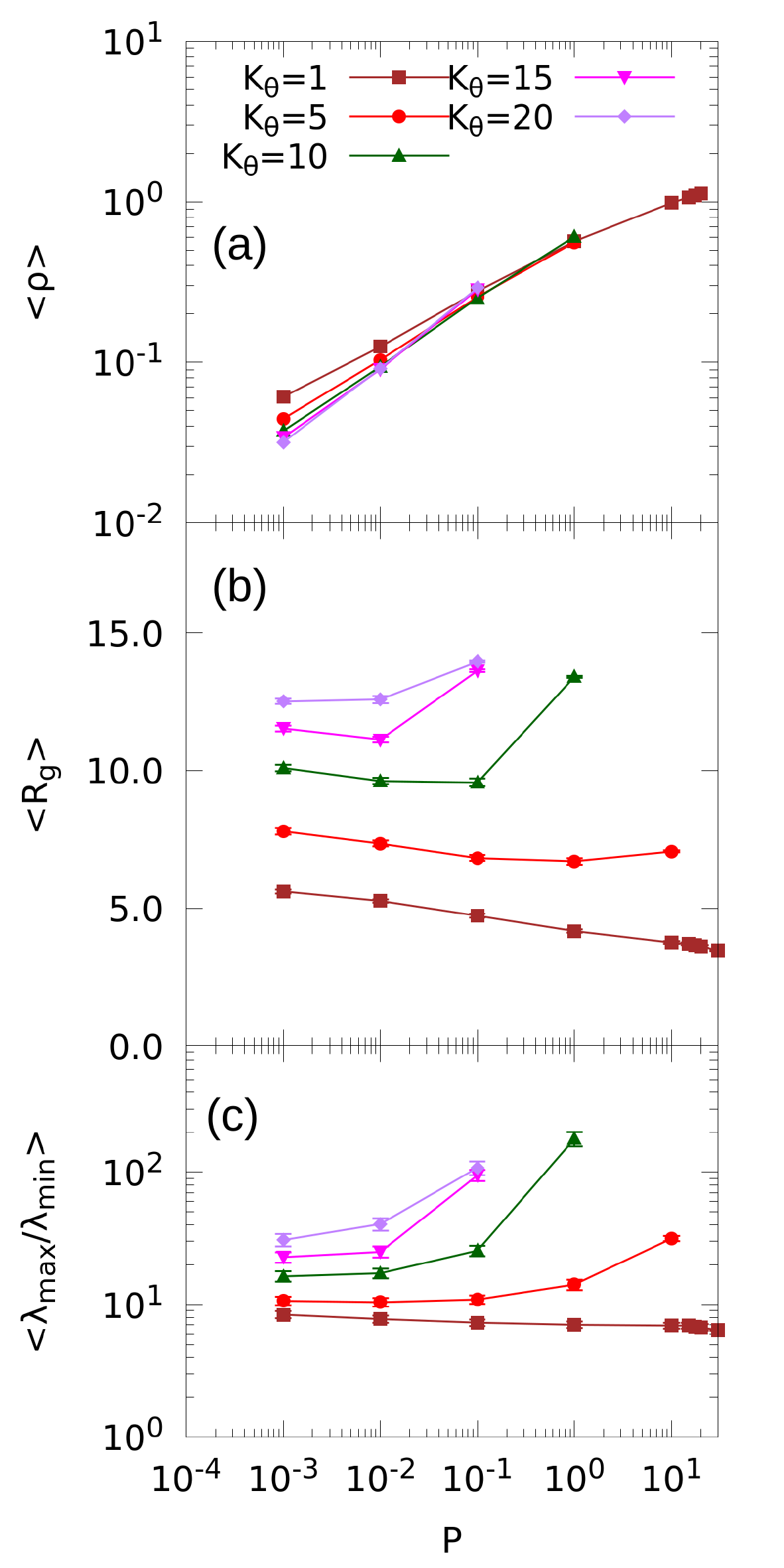}
\caption{{Average density ($\langle\rho\rangle$), radius of gyration ($\langle{R_g}\rangle$), and eigenvalue ratio ($\langle \lambda_{\text{max}}/\lambda_{\text{min}} \rangle$) as a function of applied pressure values for different $K_{\theta}$ values,  measured under equilibrium conditions.}}
\label{fig:structure}
\end{figure}

{In this section, we characterize the structural variation of the ring polymer systems across the P-$K_{\theta}$ phase-diagram to understand the possible connection between structure and emergence of slow dynamics. In Figure~\ref{fig:structure}, the variation of average density ($\langle\rho\rangle$), radius of gyration ($\langle{R_g}\rangle$), and eigenvalue ratio ($\langle \lambda_{\text{max}}/\lambda_{\text{min}} \rangle$, defined later) with pressure for different ring polymer stiffness is shown. All data is averaged over all trajectories within a time window of $0.5 \times 10^6$ within the equilibrium regime.} 

\begin{equation}
R_g = \sqrt{\lambda_1^2 + \lambda_2^2 + \lambda_3^2}
\label{eqn:rg}
\end{equation}  

{where $\lambda_1, \lambda_2, \lambda_3$ are the principle moments of inertia, calculated from the eigenvalues of the gyration tensor of each polymer. In addition, we also compute the ratio of the maximum and minimum eigenvalues to understand the shape of the ring polymers. For a globule shaped polymer, this ratio remains close to 1 but increases as the structure becomes more anisotropic. Results discussed below, for these two observables, correspond to averages over all rings within configurations sampled in the equilibrium regime.}

As can be expected, the average density of the system increases sharply with pressure for all values of  $K_{\theta}$; see Figure~\ref{fig:structure}(a). However, it should be noted that the highest values of equilibrium density achieved in our simulations, over the time window of our simulations, is for the most flexible chains (low  $K_{\theta}$). As the chain stiffness increases, the system cannot access high density diffusive regimes, but rather falls out of equilibrium within our observation time window. For example, for the highest and lowest chain stiffness considered in the study ($K_{\theta}=20$, $K_{\theta}=1$), the corresponding maximum equilibrium densities achieved are $\sim$ 0.3 and 1.13, respectively. The $\langle{R_g}\rangle$ as a function of applied pressure, exhibits varying behavior with different ring polymer stiffness seen in Figure~\ref{fig:structure}(b). For low values of $K_{\theta} \lesssim 5$, with increasing pressure, the value of $\langle{R_g}\rangle$  monotonically decreases, suggesting a more collapsed conformation adopted by the ring polymers as the density of the system goes up with applied pressure. However, as $K_{\theta}$ values increase, the $\langle{R_g}\rangle$  values change with applied pressure in a non-monotonic manner. The $\langle{R_g}\rangle$  values decrease upto a certain pressure, and then increase again, strongly suggesting that the ring polymers might be adopting more extended, plate/rod like structures at higher density (see Figure~\ref{fig:snapshot}). {The increase of shape anisotropy at large pressure for $K_{\theta} \gtrsim 5$ is reflected in the eigenvalue ratio plotted in Figure~\ref{fig:structure}(c). Because of the plate or rod like shapes, the structure of the ring polymers become more flat at larger $K_{\theta}$ values. With increasing pressure, flat structures become more favorable as it helps ring polymers to better pack in a dense environment.}

Typically, at a constant temperature, size of the molecules determines the average diffusion coefficient of a system. In case of ring polymers,  $\langle{R_g}\rangle$ values at large $K_{\theta}$ values ($ \gtrsim 5$) seem to have a non-monotonic dependence on pressure, however, the diffusion coefficient does not follow the similar trend (see Figure~\ref{fig:diffusion}). This behavior suggests that underlying mechanism of the glass formation changes as the system accesses higher P-$K_{\theta}$ state points in the phase diagram. In this region, the polymers adopt more anisotropic extended structures (see Figure~\ref{fig:single_molecule}) compared to the isotropic crumpled-like structures seen at lower values of $K_{\theta}$. Hence, theories based on diffusion of spherical objects  would no longer be applicable in this regime. Here, one has to consider the effect of entanglement to describe the dynamics of the system. The increasing effect of the threading between the ring polymers, a manifestation of entanglement, can describe the dynamical slow-down at higher $K_{\theta}$ values, which we describe in next section.

\subsection{Geometric analysis of the threading events}

\begin{figure}
\centering
\includegraphics[scale=1.2]{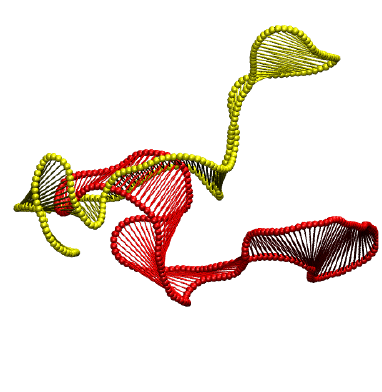}
\caption{A pictorial description of a threading event. The horizontal lines inside each ring polymer surfaces depict the minimal surfaces detected by the algorithm.}
\end{figure}

In this section, we correlate the threading of rings and emergence of slow dynamics in ring polymer systems. We develop a threading-detection algorithm, partly based on previously published triangulation algorithm ~\cite{Tsalikis_Vlassopoulos_MacroLett_2016}, and incorporating geometric analysis of individual rings {in addition}. The geometrical analysis was used to identify loops along the contour of the ring polymer which helped to calculate the area of the minimal surface with a higher accuracy as compared to the contour reduction methods~\cite{Tsalikis_Vlassopoulos_MacroLett_2016}.

To identify threading between any two ring polymers, A and B, we compute all the intersection points between bond vectors of ring polymer A and triangular planes of ring polymer B which lie on the minimal surface of ring polymer B. If number of such intersections points are greater than zero, we assume that  A is threaded into B (not the other way around). Due to geometric reasons, total number of intersections will be an even number. The quality of threading is identified by considering the length of the threaded sections~\cite{Smrek_Grosberg_MacroLett_2016, Smrek_Rosa_MacroLett_2019}.  A threading event between ring polymers A and B is defined by the two successive intersections found along the bond vectors of A.  The extent of threading, $N_{\text{th}}$ is defined as the total number of bond vectors of ring polymer A present between the two successive intersections with ring polymer B. {We denote the two sides of the minimum plane of ring B as odd-side and even-side. By construction, $N_{\text{th}}^{\text{odd}} + N_{\text{th}}^{\text{even}} = N_{\text{ring}}$, where $N_{\text{th}}^{\text{odd}}$ and $N_{\text{th}}^{\text{even}}$ denote the extent of threading at odd- and even-side, respectively.} To quantify the degree of threading, we define a function,

\begin{equation}
\Delta_{\text{th}} = 1 - \left |{\frac{2 \sum\limits_{i=\text{odd}} N_{\text{th}}^{i}}{N_{\text{ring}}} - 1} \right |
\label{eqn:threading}
\end{equation}

where $\Delta_{\text{th}}$ is the degree of threading between two ring polymers, who have at least one threading event between them. $\Delta_{th} = 1$ for $\sum_{i=odd}N^{i}_{\text{th}} = N/2$, and $\Delta_{\text{th}} = 0$ when $\sum_{i=odd}N^{i}_{\text{th}} =  0$. For a single ring polymer, the mean degree of threading, $\langle\Delta_{th}\rangle$, is determined by averaging over all $\Delta_{\text{th}}$ with all other ring polymers for all the sampled trajectories. For more details of the threading detection algorithm, see Supplementary Information, section IA.

\subsection{Threading of ring polymers and emergence of spontaneous pinning}

\begin{figure}
\centering
\includegraphics[scale=0.8]{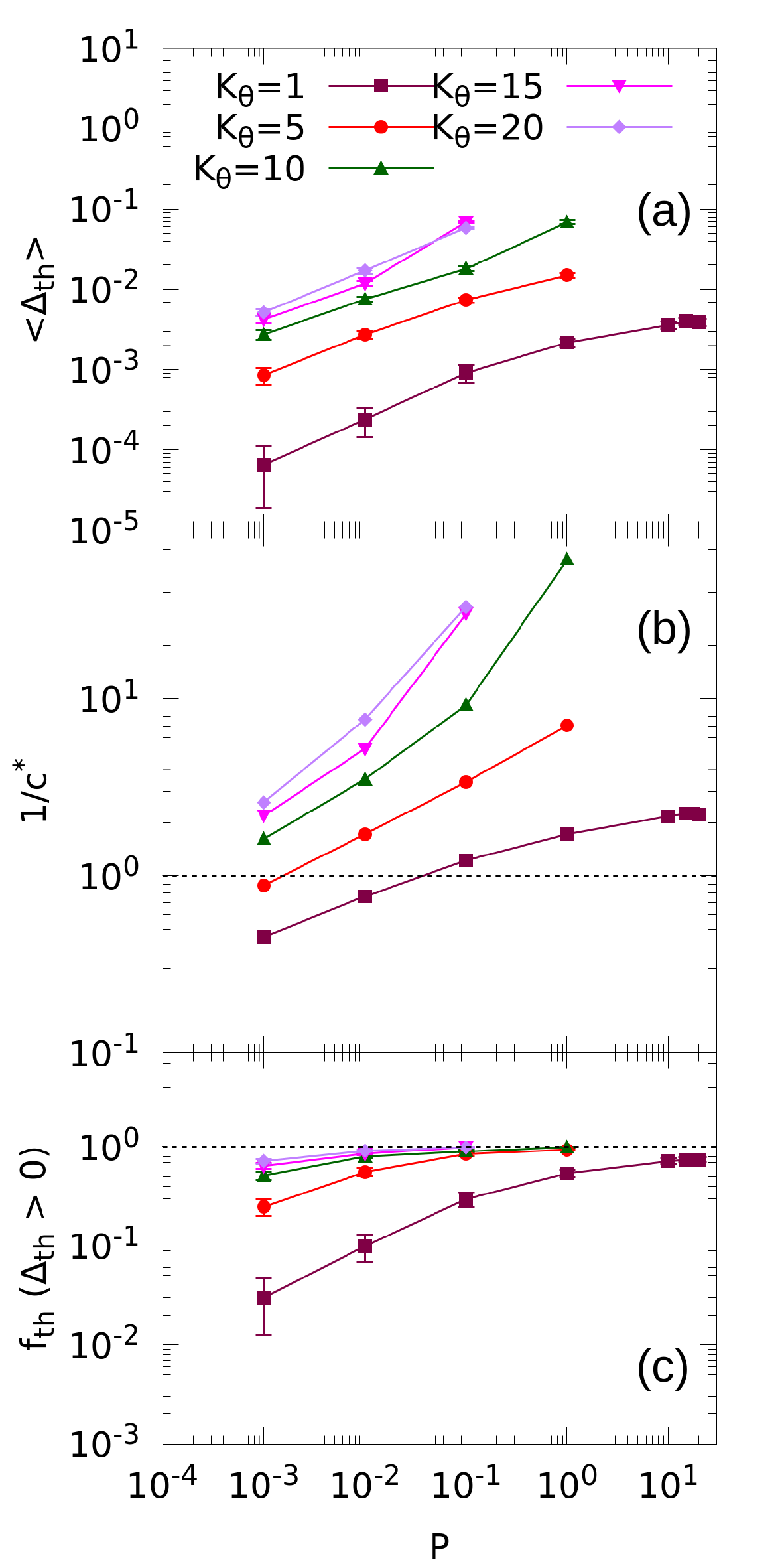}
\caption{Variation of (a) average threading factor, (b) inverse of the overlap concentration, and (c) average fraction of threaded rings with different pressure values and ring stiffness, measured within the accessible equilibrium regime. The horizontal dashed line in (b) and (c) cuts the y-axis at 1.}
\label{fig:thfac}
\end{figure}

In Figure ~\ref{fig:thfac}(a), the $\langle\Delta_{\text{th}}\rangle$ for the ring polymer systems is plotted as a function of pressure for various values of ring polymer stiffness. For all values of ring polymer stiffness, $\langle\Delta_{\text{th}}\rangle$ increases monotonically with pressure suggesting the increase in threading events with the increasing density of the ring polymer systems. While it is possible to calculate $\langle\Delta_{\text{th}}\rangle$ from the simulation trajectories, it is difficult to obtain experimentally~\cite{Gomez_Poschel_PNAS_2020}. On the other hand, $\langle{R_g}\rangle$, D, and $\langle\rho\rangle$ are readily measurable via experiments.  Here, we attempt to find some correlation between the experimentally available structural quantities and $\langle\Delta_{\text{th}}\rangle$ calculated from simulations.

 {We define an overlap concentration of the ring polymer system, $c^*$, to capture the essence of possible overlap of rings as pressure on the system is increased. The effective shape of the ring polymers can be approximated as spheres with volume $(4 \pi/3) \langle{R_g}\rangle^3$. Further, from emergent density ($\langle\rho\rangle$) due to pressure applied on the system, an effective volume per ring polymer can be defined as $\langle{V_m}\rangle= N/\langle\rho\rangle$, where $N$ is the total number of ring polymers. Using these two parameters, the overlap concentration $c^*$ is defined as:}

\begin{equation}
1/c^* = \frac{4 \pi \langle{R_g}\rangle^3}{3\langle{V_m}\rangle}
\label{eqn:effctive_size}
\end{equation}

{A higher value of  $1/c^*$ can come from either large $R_g$, suggesting possible non-collapse and anisotropic structures or smaller $V_m$ indicating threading. In Figure~\ref{fig:thfac}(a) and ~\ref{fig:thfac}(b), it can be seen that there is a strong correlation in the trend of $(1/c^*)$ and $\langle\Delta_{\text{th}}\rangle$. This correlation suggests that it is possible to estimate the average amount of threading in the ring polymer systems from the inverse of $c^*$. Overlap concentration is measured in dilute polymer solutions~\cite{Graessley_Polymer_1980} via viscosity analysis. Due a strong correlation between the threading and dynamics, it should be possible to define $c^*$ in a pure ring polymer system in a similar manner. However, an extrapolation of the data from dilute solution to dense systems is not straightforward and requires further investigation, which is beyond scope of this paper. }

\begin{figure*}[!t]
\centering
\includegraphics[scale=0.8]{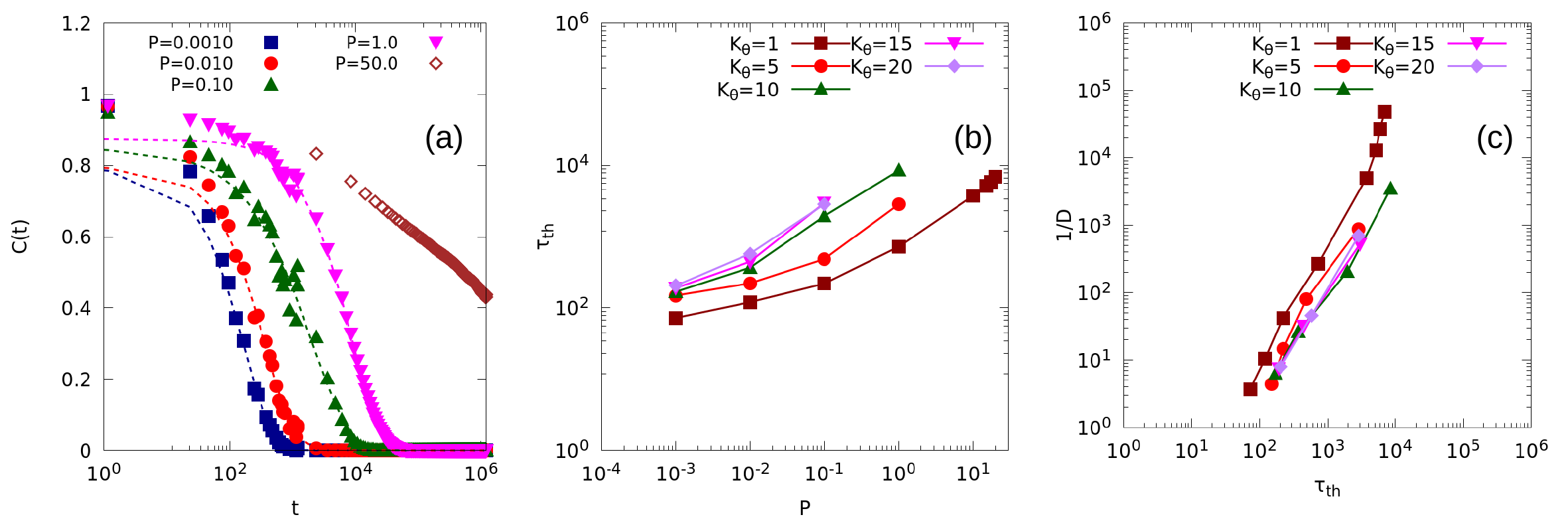}
\caption{{(a) Variation of threading correlation function $C(t)$  (see text for definition) with pressure values at $K_{\theta} = 10$. Filled symbols correspond to pressure values where equilibrium has been attained, and open symbols correspond to the case where the dynamics is not in equilibrium over the time window of observation. Lines correspond to fits to equilibrium data with stretched exponential function (see text) to obtain threading timescale  $\tau_{\text{th}}$. (b) Variation of $\tau_{\text{th}}$ with pressure, for different $K_{\theta}$. (c) Correlation plot between different timescales, i.e.  $1/D$ (a diffusive timescale) and $\tau_{\text{th}}$, for different P-$K_{\theta}$ state points.}}
\label{fig:thcorr}
\end{figure*}

{The data in Figure~\ref{fig:thfac}(b)  shows that the values of $(1/c^*)$ become much larger than 1 with increasing pressure for $K_{\theta} \gtrsim 5$. As discussed earlier, such large increase is possible only if the polymers are highly anisotropic in nature or they form highly entangled conformations. From the structural characterization of previous section, we deduce the following. For intermediate values of stiffness, $5 \lesssim K_{\theta} \lesssim 10$, the structures suggest that the anisotropy is not significant and the large value of  $(1/c^*)$  primarily can be attributed to threading. To verify this, we calculate the change in the average fraction of threaded polymers in the bulk as seen in Figure~\ref{fig:thfac}(c). We found that as the bulk system approaches the regime of increasingly slow dynamics, the fraction of threaded rings, $f_{\text{th}}$, sharply increases and saturates to 1. We have further verified that the total fraction of ring polymers participating in extensive inter-ring threading is quite high ($\Delta_{\text{th}}$ > 0.75, see Supplementary Information Figure SI.7(a)(d)). This further confirms that with increasing pressures, the dense ring polymer systems prefer a well-threaded environment. Therefore, in the glass phase, we predict that the system must contain a vast network of threaded ring polymers. For stiffer rings, i.e. $K_{\theta} > 10$, threading alone perhaps does not account for the increase in $1/c^*$ values. In this regime, the high values of $(1/c^*)$  can also be contributed by the anisotropical, plate/rod like structures at high density. Similar emergence of anisotropy in the form of cluster-glass phase for ring polymers at high $K_{\theta}$ values has also been seen previously~\cite{Slimani_Moreno_MacroLett_2014, Poier_Blaak_SoftMatter_2016, Liebetreu_Likos_ApplPolMater_2020}.}

The values of $\langle\Delta_{\text{th}}\rangle$ respond in a non-monotonic fashion with respect to ring polymer stiffness $K_{\theta}$ at some pressure values (see Supplementary Information, Figure SI.5(a)), similar to a recent work~\cite{Guo_Zhang_Polymers_2020}. This behavior can be attributed to the emergence of hollow disk phase, where rings have lower tendency to thread into each other due to higher stiffness ~\cite{Guo_Zhang_Polymers_2020}. However, a direct correlation between $\langle\Delta_{\text{th}}\rangle$ and diffusion is not straightforward. Some special form of threading between two rings, like square knots, may lead to a slow dynamics despite having a low $\Delta_{\text{th}}$ value~\cite{Soh_Doyle_PhysRevLett_2019, Chubak_Smrek_PhysRevRes_2020}. Hence, the lifetime of a threading event is more important quantity to understand the effect of threading on dynamics of the ring polymers. To understand such temporal evolution of threading event, a threading related correlation function ($C(t)$) for a pair of initially threaded
rings is calculated as, 

\begin{equation}
C(t) = \frac{\langle\Delta_{\text{th}}(t_0)\Delta_{\text{th}}(t_0+t)\rangle}{\langle\Delta_{\text{th}}^2\rangle}
\label{eqn:thcorr}
\end{equation}

{By definition, $C(t)$ varies from 1 to 0 as the threading correlation decreases between two rings.  Once un-threading of the two rings occurs, the threading event ends and thereby the computation of $C(t)$ corresponding to that pair of rings  also ends.  Any re-threading events between two rings at a later time will subsequently be treated as a new event.  Hence $C(t)$ is similar to a first passage correlator corresponding to an un-threading event between two rings that were initially entangled. This definition of $C(t)$ makes it a continuous correlation function.}  

{In Figure~\ref{fig:thcorr}(a), we show the variation of $C(t)$ at different pressures with time for a particular stiffness of the ring ($K_{\theta}=10$). We observe that, for this value of  $K_{\theta}$, in the regime of pressure where equilibrium dynamics is observed (see Fig. \ref{fig:diffusion}), $C(t)$ decreases with time and eventually decays to zero  i.e. complete decorrelation of threaded polymers occur. Further, it is evident  that such unthreading processes occur over longer timescales at larger pressures. In the pressure regime where we do not observe equilibrium diffusion dynamics within time window of measurement, $C(t)$ does not completely decorrelate, i.e, threadings continue to persist over the observation time window.}

{In the equilibrium regime, it is possible to fit the eventual decay of $C(t)$ decay with a stretched exponential function, viz. $C(t) \sim A\exp{[-(t/\tau_{\text{th}})^p]}$, where $A$ is an empirical constant, $\tau_{\text{th}}$ is the threading timescale extracted via the stretched exponential fit, and $p$ is the stretched exponent. In Figure~\ref{fig:thfac}(b), we show that  $\tau_{\text{th}}$  increases with pressure, for all the different $K_{\theta}$ values that we have explored. Together with Figure~\ref{fig:thfac}(b) and Figure~\ref{fig:thcorr}(b), this result indicates that threading leads to formation of a stable network among the ring polymers, whose lifetime increases with increasing pressure. It is also evident that for any given pressure, the persistence of threading events increases with the stiffness of the ring. } 

{A relevant question to ask is how the threading lifetimes can be correlated with the observed dynamical behavior.  In Figure~\ref{fig:thcorr}(c), we illustrate a positive correlation between $\tau_{\text{th}}$ and a diffusive timescale corresponding to the inverse of the diffusion coefficient  demonstrating that threading is a contributing factor for dynamical slowing down  with increasing pressure at all $K_{\theta}$ values. However, the extent of contribution of $\tau_{\text{th}}$ on dynamical behavior can not be extracted directly form Figure~\ref{fig:thcorr}(c). In fact, the data shows that the slowing down in diffusion happens slower than the growth in threading lifetime, with increasing pressure. Hence there are likely to be effects from other relevant relaxation processes on the dynamics of the ring polymers. Possibly, such effects are originating from different forms of inter-ring entanglements other than threading. For example, at small K, $\tau_{\text{th}}$ is much smaller and threading is not expected to be the dominant factor in the dynamical slowdown with increasing pressure. Rather, excluded volume effects due to crowding among the adjacent globule-shaped ring polymers are more contributing to its dynamics. With increasing K, rings are no longer globular and such excluded volume effect is expected to diminish but still remain relevant. However, in this regime, threading could become the more contributing factor to the dynamical slowdown with increasing stiffness, which is realized by order of magnitude increase of the $\tau_{\text{th}}$ data with pressure in Figure~\ref{fig:thcorr}(c).}

\section{Conclusion and perspectives}

{Emergence of dynamically arrested states in topologically constrained polymer systems is an exciting area which can serve as model platform to explore the dynamics of deformable entities and emergent new phases. Earlier simulations have demonstrated the occurrence of glassy dynamics in ring polymers via artificial pinning of the ring polymers, motivated by similar explorations of random pinning in supercooled atomistic liquids \cite{ozawa2015equilibrium}. It is not very clear what is the role played by both the ring polymer stiffness and ambient  pressure conditions in the formation of such novel phases and whether the presence of quenched disorder is necessarily required for facilitating the path towards the formation of topological glasses.}

{In this work, we explore the phase space spanned by ring polymer stiffness and ambient pressure  to elucidate the emergence of glassy states via large scale simulations of a thermal assembly of ring polymers formed by hundred monomers. Our work evidences that large ring size is not a prerequisite for observing threading in the equilibrated systems of semi-stiff rings. We show that dynamical slowing down and  thereby the tendency towards glassiness emerges as we scan across the phase space. We demonstrate that stiffer rings are likely to get dynamically arrested at smaller ambient pressures, and correspondingly smaller densities.} By calculating various structural properties of ring polymers in the P-$K_\theta$, we show that the ring polymers show a variety of conformations from completely crumpled structures (flexible ring polymers) to elongated and stiff plate like conformations (stiff ring polymers). These structural results strongly suggest that the ring polymers considered here encounter significant dynamical constraints and thus perhaps achieve better close packing via clustering and shape modifications, some of which have been see earlier ~\cite{Bernabei_Likos_SoftMatter_2013, Poier_Blaak_Macromol_2015, Poier_Blaak_Macromol_2015, Poier_Blaak_SoftMatter_2016, Liebetreu_Likos_ApplPolMater_2020} . As shown in Figure~\ref{fig:single_molecule}, at low $K_{\theta}$, the system becomes more closed and globular with increasing pressure. At the medium $K_{\theta}$ range, the rings take a rod like shape at high pressure. At large $K_{\theta}$, rings take a circular shape which minimizes its energy and helps packing via clustering.

To better understand the emergent slow dynamics with varying ring stiffness and pressure, we explored the possibility of threading among the ring polymers, which has been proposed as a mechanism for the formation of the topological glass that is unique to these objects. For identifying such  conformations, we develop an improved threading algorithm with specific inputs from the geometry of the ring polymers which gives an excellent quantification of the extent of threading. Our threading analyses  strongly suggests that for any specific ring stiffness with increasing pressure values, the average threading factor increases and contributes to the slow dynamics. Further, the occurrence of such threading events increase with increasing ring stiffness, at any particular ambient pressure condition. {We define an {\it overlap concentration} for the polymer systems -- usually measured experimentally in dilute polymer solutions~\cite{Graessley_Polymer_1980} -- which strongly correlates with the average threading factor measured from the simulation. This result has the potential to promote experimental studies on measurement of threading in dense ring polymer systems.} We also measure the life time of a threaded event and find that as the pressure increases, the ring polymers show more long-lived threaded events, in sync with increasing slow down in the dynamics.  Taken together all the structural and dynamical properties measured in our simulations,  {we clearly demonstrate the spontaneous emergence of slow, glassy dynamics in ring polymer systems -- by suitable} selection of ring polymer stiffness and pressures applied, slow dynamics can be induced in ring polymers at very low pressure values. We also note that we are observing the formation of such topologically constrained states for relatively smaller ring sizes compared to what has been explored earlier, indicating that the access to such states can be obtained by tuning the properties of the ring.   {We hope that our work will motivate experimental studies to validate our findings.}

In future, interesting avenues of study could include the possibility of introducing crowder particles into the ring polymer systems or adding differential activities~\cite{smrek2020active, Chubak_Smrek_PhysRevRes_2020,Chubak_Smrek_arxiv_2021,Vlassopoulos_JourClubJPCM_2021} to the ring polymer via temperature changes to further probe glass formation in such systems.

\section*{Author Contributions}
PC and SV conceived the study. PR carried out the simulations. All authors contributed to writing of the manuscript.

\section*{Conflicts of interest}
There are no conflicts to declare.

\section*{Acknowledgments}
The authors thank the HPC facility at IMSc, Chennai, for providing the necessary computational platform for this research.



\balance


\bibliographystyle{rsc} 
\providecommand*{\mcitethebibliography}{\thebibliography}
\csname @ifundefined\endcsname{endmcitethebibliography}
{\let\endmcitethebibliography\endthebibliography}{}

\end{document}